\documentclass[aps,prd,superscriptaddress,nofootinbib,amsmath,amsfonts,preprintnumbers,notitlepage,10pt,english]{revtex4}
\usepackage{amsmath}
\usepackage{amssymb}
\usepackage{babel}

\makeatletter

\@ifundefined{textcolor}{}
{%
 \definecolor{BLACK}{gray}{0}
 \definecolor{WHITE}{gray}{1}
 \definecolor{RED}{rgb}{1,0,0}
 \definecolor{GREEN}{rgb}{0,1,0}
 \definecolor{BLUE}{rgb}{0,0,1}
 \definecolor{CYAN}{cmyk}{1,0,0,0}
 \definecolor{MAGENTA}{cmyk}{0,1,0,0}
 \definecolor{YELLOW}{cmyk}{0,0,1,0}
 }

\@ifundefined{textcolor}{}{%
 \definecolor{BLACK}{gray}{0}
 \definecolor{WHITE}{gray}{1}
 \definecolor{RED}{rgb}{1,0,0}
 \definecolor{GREEN}{rgb}{0,1,0}
 \definecolor{BLUE}{rgb}{0,0,1}
 \definecolor{CYAN}{cmyk}{1,0,0,0}
 \definecolor{MAGENTA}{cmyk}{0,1,0,0}
 \definecolor{YELLOW}{cmyk}{0,0,1,0}
 }

\makeatother

\begin{document}

\preprint{RESCEU-27/11}

\title{Black hole perturbation in parity violating gravitational theories}

\author{Hayato Motohashi}
\affiliation{Department of Physics, Graduate School of Science,
The University of Tokyo, Tokyo 113-0033, Japan}
\affiliation{Research Center for the Early Universe (RESCEU),
Graduate School of Science, The University of Tokyo, Tokyo 113-0033, Japan}

\author{Teruaki Suyama}
\affiliation{Research Center for the Early Universe (RESCEU),
Graduate School of Science, The University of Tokyo, Tokyo 113-0033, Japan}

\begin{abstract}
We study linear perturbations around static, spherically-symmetric spacetimes in $f(R,C)$ gravitational theories whose Lagrangians depend on the Ricci scalar $R$ and the parity violating Chern-Simons term $C$. 
By an explicit construction, we show that the Hamiltonian for the perturbation variables is not bounded from below, suggesting that such a background spacetime is unstable against perturbations. 
This gives a strong limit on a phenomenological gravitational model which violates parity. 
We also show that either $R={\rm const}$ or $\frac{\partial^2 f}{\partial R \partial C}=0$ is a necessary and sufficient condition for the stability. 
We then implement in detail the perturbation analysis for such theories which satisfy the stability conditions. 
For $\ell \ge 2$, where $\ell$ is the usual integer for the multipole expansion, the number of propagating modes is three, one from the odd and the other two from the even, all of which propagate at the speed of light.
Unlike in the case of $f(R)$ theories, these modes are coupled to each other, which can be used as a distinctive feature to test the parity violating theories from observations.
The no-ghost conditions and no-tachyon conditions are the same as those in $f(R)$ theories.
For the dipole perturbations, the odd and the even modes completely decouple.
The odd mode gives a slowly rotating black hole solution whose metric is linearized in its angular momentum.
We provide an integral expression of this solution.
On the other hand, the even mode propagates at the speed of light.
For the monopole perturbation, in addition to a mode which simply shifts the mass of the background black hole, there also exists one even mode that propagates at the speed of light.

\end{abstract}

\date{\today}

\maketitle

\section{Introduction}
General relativity (GR) has been frequently tested both experimentally and observationally over many decades to check whether it is really the correct theory of gravity \cite{Will}.
In the weak gravitational field regime, like in the solar system or on the Earth, no deviations from GR have been detected so far.
In this sense, GR is a successful theory of gravity.
However, the need for the introduction of dark energy to explain accelerated expansion of the universe (for recent reviews on dark energy, see e.g., Refs.~\cite{Copeland:2006wr, amendola}) may be a signal that GR breaks down in a regime where the strong field effects become important.
Furthermore, although there are few observational tests of GR in the strong field regime, in the future, we will be able to probe the nature of strong gravity,
for example, by observing the gravitational waves coming from the vicinity of black holes (BHs) \cite{Yagi:2009zm,Yagi:2009zz}.
These facts have provoked alternative theories of gravity such as $f(R)$ theories (for recent reviews of $f(R)$ theories, see e.g. Refs.~\cite{Sotiriou:2008rp,DeFelice:2010aj}) and have led us to understand theoretically what kinds of different phenomena are expected in such theories \cite{Starobinsky:1980te,Capozziello:2002rd,Carroll:2003wy,Bean:2006up,Carloni:2007yv,Appleby:2007vb,Starobinsky:2007hu,Motohashi:2009qn,Motohashi:2010tb,Motohashi:2010sj,Motohashi:2011wn,Motohashi:2011wy}.

In light of this situation, it is interesting to consider gravitational theories which violate parity due to the so-called Chern-Simons(CS) term, or the Pontryagin density, $C \equiv \frac{1}{2} \epsilon_{\alpha \beta \gamma \delta} R^{\alpha \beta}_{~~~\mu \nu} R^{\gamma \delta \mu \nu}$, where $\epsilon_{\alpha \beta \gamma \delta}$ is the totally antisymmetric tensor.
The presence of the $\epsilon_{\alpha \beta \gamma \delta}$ tensor manifests the parity violation.
Because the CS term can be expressed as a divergence, simply adding $C$ into the Einstein-Hilbert action amounts to an addition of a total derivative term in the action and does not change the theory.
A gravitational theory in which the CS term is coupled to an external scalar function was introduced in Ref.~\cite{Jackiw:2003pm}.
The external scalar function was promoted to a dynamical field in Ref.~\cite{Smith:2007jm} and many studies based on such theories have appeared \cite{Yunes:2010yf,Pani:2011xj,Cambiaso:2010un,Garfinkle:2010zx,Molina:2010fb,Yunes:2009ch,Sopuerta:2009iy,Yunes:2009hc,Konno:2009kg,Yunes:2008ua,Konno:2008np,Yunes:2007ss,Grumiller:2007rv,Konno:2007ze}. 
For a recent review on the Chern-Simons gravity, see Ref.~\cite{Alexander:2009tp}.
Noticeable features of the CS term are that it identically vanishes for the Friedmann-Lema\^\i
tre-Robertson-Walker (FLRW) metric and even for the scalar type perturbations on top of it, and for the spherically symmetric metric.
As a result, the cosmological and the Solar System constraints achieved so far 
constrain the CS gravity only very loosely.
Therefore, we need to go beyond such simple spacetimes to make the lurking CS term yield different phenomena from GR.
For example, the CS term enters the game in the tensor perturbations on the FLRW background, which suggests that observing gravitational waves is an effective approach to test the CS-gravity \cite{Soda:2011am,Satoh:2010ep,Nishizawa:2009bf,Seto:2008sr,Seto:2007tn,Satoh:2007gn,Saito:2007kt}. 
Another useful approach is to consider perturbations around the static, spherically-symmetric background such as Schwarzshild BH, where the CS effects are expected to show up \cite{Yunes:2010yf,Pani:2011xj,Cambiaso:2010un,Garfinkle:2010zx,Sopuerta:2009iy,Yunes:2009hc,Konno:2009kg,Yunes:2007ss}.

In this paper, we consider the gravitational theories whose Lagrangian is a general function of $R$ and $C$, $f(R,C)$, and develop linear perturbation theory around the static and spherically symmetric spacetime.
Similar analysis has been done for $f(R,{\cal G})$ (${\cal G}$ is the Gauss-Bonnet term) theories in Ref.~\cite{DeFelice:2011ka}.
Unlike in the case of $f(R)$ theories which can be mapped into equivalent
theories where a scalar field having self-interacting potential is minimally coupled to Einstein-Hilbert gravity, $f(R,C)$ theories cannot be mapped into theories where $C$ is coupled solely to a dynamical scalar field due to nontrivial transformation property of the CS term under the conformal transformation.
Our aim is to clarify both quantitative and qualitative behaviors of the perturbations.
To be more precise, we will derive no-ghost and no-tachyon conditions which are necessary to ensure stability of the background spacetime against perturbation,  obtain dispersion relations for the propagating modes and find features that are characteristic to parity violating theories.

The BH perturbation for the so-called nondynamical Chern-Simons theories where 
the Lagrangian consists of the Einstein-Hilbert term plus the CS term $C$ multiplied by a nondynamical scalar field has been studied in Ref.~\cite{Yunes:2007ss}.
As we will show in the next section, the general $f(R,C)$ theories we consider do not cover such nondynamical CS theories. 
Correspondingly, the results presented in Ref.~\cite{Yunes:2007ss} cannot be applied to our study.

The organization of this paper is as follows.
In Sec.~\ref{sec-back}, we derive the gravitational field equations and apply them to the static and spherically symmetric background.
In Sec.~\ref{sec-pert}, we will develop linear perturbation analysis on that background by expanding the action in second order perturbations and show that the perturbations are unstable in general.
Then in Sec.~\ref{sec-sp}, we will study in detail a special class of theories in which the instability of the perturbations is avoided.
The last section is the conclusion.

\section{Background Equations}
\label{sec-back}
We study $f(R,C)$ theory, where the action is described by a general function of Ricci scalar $R$ and CS term $C \equiv \frac{1}{2} \epsilon_{\alpha \beta \gamma \delta} R^{\alpha \beta}_{~~~\mu \nu} R^{\gamma \delta \mu \nu}$,
\begin{equation}
S=\frac{M_P^2}{2} \int d^4x~\sqrt{-g} f(R,C). \label{act1}
\end{equation}
Here, $M_P=1/\sqrt{8\pi G_N} \simeq 4.34 \times 10^{-6} {\rm g}$ is the reduced Planck mass.
We can rewrite the action (\ref{act1}) as
\begin{equation}
S=\frac{M_P^2}{2} \int d^4x~\sqrt{-g} \left( RF(\lambda,s)+W(\lambda,s)C-V(\lambda,s) \right), \label{act2}
\end{equation}
where $\lambda$ and $s$ are auxiliary fields and 
\begin{equation}
F(\lambda,s)=\frac{\partial f(\lambda,s)}{\partial \lambda},~~~W(\lambda,s)=\frac{\partial f(\lambda,s)}{\partial s},~~~V(\lambda,s)=\lambda F(\lambda,s)+s W(\lambda,s)-f(\lambda,s). \label{def}
\end{equation}
The variations with respect to $\lambda$ and $s$ yield the following constraints:
\begin{equation}
\lambda=R,~~~~~s=C.
\end{equation}
By substituting them, it can be confirmed that the second action (\ref{act2}) actually reduces to the first one (\ref{act1}).
The gravitational field equations obtained from the action (\ref{act2}) are given by
\begin{equation}
X_{\mu \nu}=R_{\mu \nu}-\frac{1}{2} R g_{\mu \nu}-\frac{1}{F} \bigg[ \nabla_\mu \nabla_\nu F-g_{\mu \nu} \Box F-2 \epsilon^{\alpha \beta \gamma}_{~~~~(\mu} \left( R^\sigma_{~\nu) \alpha \beta} \nabla_\sigma \nabla_\gamma W +2 \nabla_\gamma W ~\nabla_\alpha R_{\nu) \beta} \right)-\frac{1}{2} g_{\mu \nu}V \bigg]=0.
\end{equation}
Here, the parentheses around the indices indicate symmetrization, $S_{(\mu\nu)}\equiv (S_{\mu\nu}+S_{\nu\mu})/2$.

Let us show that the nondynamical CS theories considered in Ref.~\cite{Yunes:2007ss},
whose action is given by
\begin{equation}
S_{\rm non-dynamical}=\frac{M_P^2}{2} \int d^4x~\sqrt{-g} \left( R+W C \right), \label{non-dynamical}
\end{equation}
are not covered by Eq.~(\ref{act2}).
In Eq.~(\ref{non-dynamical}), $W$ is treated as a field. Apparently, we can reproduce Eq.~(\ref{non-dynamical}) by setting $F=1$ and $V=0$ in Eq.~(\ref{act2}). However, by using the definitions of $F,~W$ and $V$ given by Eqs.~(\ref{def}), the conditions $F=1$ and $V=0$ lead to $W$ as a constant. Hence, the action describes GR instead of Eq.~(\ref{non-dynamical}). Thus, the nondynamical CS theories are not included in $f(R,C)$ theory, and vice versa.
Clearly, dynamical CS theory, whose Lagrangian has the kinetic and potential terms in addition to Eq.~(\ref{non-dynamical}), is also not included.
The analysis for nondynamical and dynamical CS theory is given in Ref.~\cite{Motohashi:2011ds}.

Throughout the paper, we consider a static and spherically symmetric spacetime as a background,
{\it i.e.}, we set the background metric as
\begin{equation}
ds^2=-A(r) dt^2+\frac{dr^2}{B(r)}+r^2 \left( d\theta^2+\sin^2 \theta ~d\varphi^2 \right).
\end{equation}
For this metric, nonvanishing and independent elements of $X_{\mu \nu}$ are given by
\begin{eqnarray}
X_{tt}&=&-\frac{A \left(r \left(r B' F'+2 r B F''+4 B F'+r V\right)+2 F \left(r B'+B-1\right)\right)}{2 r^2 F}, \\
X_{rr}&=&\frac{r B A' \left(r F'+2 F\right)+A \left(r \left(4 B F'+r V\right)+2 (B-1) F\right)}{2 r^2 A B F}, \\
X_{\theta \theta}&=&\frac{r \left(-rBF A'^2+A \left(rF A' B'+2B \left(r A' F'+F \left(r A''+A' \right)\right)\right)+2 A^2 \left(r B' F'+F B'+2r B F''+2B F'+rV\right)\right)}{4A^2 F},
\end{eqnarray}
where $'$ means derivative with respect to $r$.
The function $W$ does not appear in $X_{\mu \nu}$ because of the spherical symmetry of the background.
We can convert the background equations $X_{tt}=0,~X_{rr}=0$ and $X_{\theta \theta}=0$ to
express $A'',~F''$ and $V$ in terms of $A,B$ and $F$ and their first derivatives,
\begin{eqnarray}
A''&=&\frac{-r A A' \left(r F B'+2 B\left(r F'+F\right)\right)+r^2 B FA'^2+2 A^2 \left(F \left(r B'+2 B-2\right)+2 r B F'\right)}{2 r^2AB F}, \\
F''&=&\frac{\left(r F'+2 F\right) \left(B A'-A B'\right)}{2 r A B}, \\
V&=&-\frac{r BA' \left(r F'+2 F\right)+2 A \left(2 r B F'+(B-1) F\right)}{r^2 A}.
\end{eqnarray}
We use these equations when the second derivative terms appear as a result of integration by parts to rearrange terms in the second order action for the perturbation.

\section{Perturbation}
\label{sec-pert}
In this section, we calculate the perturbative action. We start from the metric perturbation with a brief review of the Regge-Wheeler formalism. As a result of the parity violation term, the odd and even modes do not decouple from each other within the system of equations and we have to deal with it all together. 
We derive the second order action and find that there exists an instability due to the presence of the higher derivative terms. 
We assume $\ell \ge 1$ when we construct our final action after the elimination of auxiliary fields.
There is also an issue regarding residual gauge degrees of freedom for $\ell=0,1$ and we will consider those modes separately in the next section.

\subsection{Decomposition of the metric perturbation}
Before studying the metric perturbation of a spherically symmetric static spacetime for $f(R,C)$ theories, let us briefly review the formalism developed by Regge and Wheeler \cite{Regge:1957td}, and Zerilli \cite{Zerilli:1970se} to decompose the metric perturbations according to their transformation properties under two-dimensional rotations. 
Although Regge, Wheeler and Zerilli considered the perturbation of the Schwarzschild spacetime (namely in GR), the formalism solely relies on the properties of spherical symmetry and can be applied to $f(R,C)$ theories as well.

Let us denote the metric, slightly perturbed from a spherically symmetric
static spacetime, by $g_{\mu\nu}=g_{\mu\nu}^{0}+h_{\mu\nu}$. Hence $h_{\mu\nu}$ represents infinitesimal quantities. Then, under two-dimensional rotations on a sphere, $h_{tt},h_{tr}$ and $h_{rr}$ transform as scalars, $h_{ta}$ and $h_{ra}$ transform as vectors and $h_{ab}$ transforms as a tensor (subscripts $a,b$ are either $\theta$ or $\varphi$). 
First, any scalar $s$  can be decomposed into the sum of spherical harmonics as
\begin{equation}
s(t,r,\theta,\varphi)=\sum_{\ell,m}s_{\ell m}(t,r)Y_{\ell m}(\theta,\varphi).\label{scalar-decomposition}
\end{equation}
Second, any vector $V_{a}$ can be decomposed into a divergence part and a divergence-free part as follows:
\begin{equation}
V_{a}(t,r,\theta,\varphi)=\nabla_{a}\Phi_{1}+E_{a}^b\nabla_b\Phi_{2},
\end{equation}
where $\Phi_{1}$ and $\Phi_{2}$ are scalars and $E_{ab}\equiv\sqrt{\det\gamma}~\epsilon_{ab}$
with $\gamma_{ab}$ being the two-dimensional metric on the sphere and $\epsilon_{ab}$ being the totally antisymmetric symbol with $\epsilon_{\theta \varphi}=1$. 
Here $\nabla_{a}$ represents the covariant derivative with respect to the metric $\gamma_{ab}$.
Since $V_{a}$ is a two-component vector, it is completely specified by the potentials $\Phi_{1}$ and $\Phi_{2}$. 
We can then apply the scalar decomposition (\ref{scalar-decomposition}) to $\Phi_{1}$ and $\Phi_{2}$ to decompose the vector quantity $V_a$ into spherical harmonics.

Finally, any symmetric tensor $T_{ab}$ can be decomposed as
\begin{equation}
T_{ab}(t,r,\theta,\varphi)=\nabla_{a}\nabla_{b}\Psi_{1}+\gamma_{ab}\Psi_{2}+\frac{1}{2}\left(E_{a}{}^{c}\nabla_{c}\nabla_{b}\Psi_{3}+E_{b}{}^{c}\nabla_{c}\nabla_{a}\Psi_{3}\right),
\end{equation}
where $\Psi_{1},~\Psi_{2}$ and $\Psi_{3}$ are scalars. Since $T_{ab}$ has three
independent components, $\Psi_{1},~\Psi_{2}$ and $\Psi_{3}$ completely
specify $T_{ab}$. Then we can again apply the scalar decomposition
(\ref{scalar-decomposition}) to $\Psi_{1},~\Psi_{2}$ and $\Psi_{3}$
to decompose the tensor quantity into spherical harmonics. We refer to the variables accompanied by $E_{ab}$ as odd-type variables and the others as even-type variables.

Let us now apply this decomposition to the metric perturbation $h_{\mu \nu}=g_{\mu \nu}-g^{(0)}_{\mu \nu}$.
For the odd-type perturbations, they can be written as
\begin{eqnarray}
 &  & h_{tt}=0,~~~h_{tr}=0,~~~h_{rr}=0,\\
 &  & h_{ta}=\sum_{\ell, m}h_{0,\ell m}(t,r)E_{ab}\partial^{b}Y_{\ell m}(\theta,\varphi),\\
 &  & h_{ra}=\sum_{\ell, m}h_{1,\ell m}(t,r)E_{ab}\partial^{b}Y_{\ell m}(\theta,\varphi),\\
 &  & h_{ab}=\frac{1}{2}\sum_{\ell, m}h_{2,\ell m}(t,r)\left[E_{a}^{~c}\nabla_{c}\nabla_{b}Y_{\ell m}(\theta,\varphi)+E_{b}^{~c}\nabla_{c}\nabla_{a}Y_{\ell m}(\theta,\varphi)\right]. \label{odd-ab}
\end{eqnarray}
For the even-type perturbations, they can be written as
\begin{eqnarray}
 &  & h_{tt}=A(r)\sum_{\ell, m}H_{0,\ell m}(t,r)Y_{\ell m}(\theta,\varphi),\\
 &  & h_{tr}=\sum_{\ell, m}H_{1,\ell m}(t,r)Y_{\ell m}(\theta,\varphi),\\
 &  & h_{rr}=\frac{1}{B(r)}\sum_{\ell, m}H_{2,\ell m}(t,r)Y_{\ell m}(\theta,\varphi),\\
 &  & h_{ta}=\sum_{\ell, m}\beta_{\ell m}(t,r)\partial_{a}Y_{\ell m}(\theta,\varphi),\\
 &  & h_{ra}=\sum_{\ell, m}\alpha_{\ell m}(t,r)\partial_{a}Y_{\ell m}(\theta,\varphi),\\ 
 &  & h_{ab}=\sum_{\ell, m} K_{\ell m}(t,r) g_{ab} Y_{\ell m}(\theta,\varphi)+\sum_{\ell, m} G_{\ell m}(t,r) \nabla_a \nabla_b Y_{\ell m}(\theta,\varphi)\,. \label{hab}
\end{eqnarray}
Because of general covariance, not all the metric perturbations are physical 
in the sense that some of them can be set to vanish by using the gauge 
transformation $x^{\mu}\to x^{\mu}+\xi^{\mu}$, where $\xi^{\mu}$ is infinitesimal function.
A gauge transformation acting on the odd-type perturbations can be written as
\begin{equation}
\xi_a =\sum_{\ell, m} \Lambda_{\ell m}(t,r) E_a^{~b} \partial_b Y_{\ell m}(\theta,\varphi),
\end{equation}
where $\Lambda_{\ell m}(t,r)$ are arbitrary functions.
Gauge transformations on even-type perturbations can be written as
\begin{equation}
\xi_t=\sum_{\ell, m} T_{\ell m}(t,r)Y_{\ell m}(\theta,\varphi),~~~\xi_r=\sum_{\ell, m} R_{\ell m}(t,r)Y_{\ell m}(\theta,\varphi),~~~\xi_a=\sum_{\ell, m} \Theta_{\ell m}(t,r) \partial_a Y_{\ell m}(\theta,\varphi),
\end{equation}
where $T_{\ell m}(t,r),~T_{\ell m}(t,r)$ and $\Theta_{\ell m}(t,r)$ are arbitrary functions.
Under these gauge transformations, we can find the transformation rule for each metric component. 
For the odd-type perturbations, we find
\begin{eqnarray}
&&h_{0,\ell m}(t,r) \to h_{0,\ell m}(t,r)+{\dot \Lambda_{\ell m}}(t,r), \label{h0-gauge}\\
&&h_{1,\ell m}(t,r) \to h_{1,\ell m}(t,r)+\Lambda_{\ell m}'(t,r)-\frac{2}{r} \Lambda_{\ell m}(t,r), \label{h1-gauge}\\
&&h_{2,\ell m}(t,r) \to h_{2,\ell m}(t,r)+2 \Lambda_{\ell m}(t,r),
\end{eqnarray}
where a dot denotes the time derivative.
For the even-type perturbations, we find 
\begin{eqnarray}
&&H_{0,\ell m}(t,r) \to H_{0,\ell m}(t,r)+\frac{2}{A} {\dot T_{\ell m}}(t,r)-\frac{A'B}{A} R_{\ell m}(t,r), \\
&&H_{1,\ell m}(t,r) \to H_{1,\ell m}(t,r)+{\dot R_{\ell m}}(t,r)+T_{\ell m}'(t,r)-\frac{A'}{A}T_{\ell m}(t,r), \\
&&H_{2,\ell m}(t,r) \to H_{2,\ell m}(t,r)+2BR_{\ell m}'(t,r)+B'R_{\ell m}(t,r), \\
&&\beta_{\ell m}(t,r) \to \beta_{\ell m}(t,r)+T_{\ell m}(t,r)+{\dot \Theta_{\ell m}}(t,r), \\
&&\alpha_{\ell m}(t,r) \to \alpha_{\ell m}(t,r)+R_{\ell m}(t,r)+\Theta_{\ell m}'(t,r)-\frac{2}{r} \Theta_{\ell m}(t,r), \\
&&K_{\ell m}(t,r) \to K_{\ell m}(t,r)+\frac{2B}{r} R_{\ell m}(t,r), \\
&&G_{\ell m}(t,r) \to G_{\ell m}(t,r)+2 \Theta_{\ell m}(t,r).
\end{eqnarray}

Since no derivative appears in the transformation rule for $h_2$, we can completely fix a gauge by imposing a condition $h_2=0$.
This gauge fixing is called Regge-Wheeler gauge.
For the even-type perturbations, complete gauge fixing is achieved by imposing $\beta=0,~K=0$ and $G=0$. 
We will use these gauge conditions in the calculation of the second order action.

We remark here that the parity violation induces the coupling between odd and the even modes.
This can be understood by comparing $f(R,C)$ gravity with its special case, $f(R)$ gravity.
In $f(R)$ gravity, where $f$ depends only on $R$ and the theory does not violate parity, the linearized equations of motion (or equivalently, the second order action) for $h_{\mu\nu}$ can be decomposed into ones that only contain odd-type perturbations and ones that only contain even-type ones.
This decoupling drastically simplifies the perturbation analysis since it allows us to consider equations (or an action) with fewer perturbation variables.
However, this decomposition does not hold for $f(R,C)$ gravity due to the explicit violation of parity.
Therefore, we must treat both the odd and the even modes at the same time,
which we will do in the following sections.

\subsection{Second order action}
In addition to the metric perturbations, we also need to perturb the other functions $\lambda$ and $s$ that appear in the action.
Just for later convenience, instead of perturbing $\lambda$ and $s$ as the fundamental fields,
we treat $\delta F$ and $\delta W$ as perturbation variables.
These fields must be also decomposed into the spherical harmonics,
\begin{equation}
\delta F=\sum_{\ell, m}\delta F_{\ell m}(t,r)Y_{\ell m}(\theta,\varphi),~~~~~\delta W=\sum_{\ell, m}\delta W_{\ell m}(t,r)Y_{\ell m}(\theta,\varphi).
\end{equation}
The relation between ($\delta F$,~$\delta W$) and ($\delta \lambda$,~$\delta s$) is given by
\begin{equation}
\delta F=F_\lambda \delta \lambda+F_s \delta s,~~~~~\delta W=F_s \delta \lambda+W_s \delta s,
\end{equation}
where we have used the identity $F_s=W_\lambda=\frac{\partial^2 f(\lambda,s)}{\partial \lambda \partial s}$ to replace $W_\lambda$ with $F_s$ in the second equation.

With these perturbation variables, expanding the action (\ref{act2}) to second order (the first order part automatically vanishes because of the background equations) yields the following action,
\begin{eqnarray}
S=\int dt~dr~{\cal L},
\end{eqnarray}
where ${\cal L}$ is written as
\begin{eqnarray}
{\cal L}&=&H_0 (a_1 H_2+a_2 H_2'+a_3 \alpha+a_4 \alpha'+a_5 \delta F+a_6 \delta F'+a_7 \delta F''+a_8 h_0+a_9 h_0'+a_{10} h_0''+a_{11} {\dot h_1}+a_{12} {\dot h_1}') \nonumber \\
&&+b_1 H_1^2+H_1(b_2 {\dot {\delta F}}'+b_3 {\dot {\delta F}}+b_4 {\dot H_2}+b_5 {\dot \alpha}+b_6 {\dot h_0}+b_7 {\dot h_0}'+b_8 h_1+b_9 {\ddot h_1}) \nonumber \\
&&+c_1 H_2^2+H_2 (c_2 \alpha+c_3 \delta F'+c_4 \delta F+c_5 h_0'+c_6 h_0+c_7 {\dot h_1})+c_8 {\dot H_2} {\dot {\delta F}} \nonumber \\
&&+d_1 {\dot \alpha}^2+d_2 \alpha^2+\alpha (d_3 \delta F'+d_4 \delta F+d_5 {\ddot h_0}'+d_6 {\ddot h_0}+d_7 h_0'+d_8 h_0+d_9 h_{1,ttt}+d_{10}{\dot h_1} \nonumber \\
&&+e_1 h_0^2+h_0 (e_2 {\dot h_1}+e_3 \delta W'+e_4 \delta W)+e_5 h_0' {\dot h_1}+e_6 h_0'^2+e_7 {\dot h_1}^2+e_8 h_1^2+e_9 {\dot h_1} \delta W \nonumber \\
&&+f_1 \delta F^2+f_2 \delta F \delta W+f_3 \delta W^2. \label{bare-lag}
\end{eqnarray}
Since different $(\ell,~m)$ modes do not mix with each other, we pick up particular $(\ell,~m)$ modes.
Because of the spherical symmetry of the background spacetime, the action for $m \neq 0$ modes takes exactly the same form as that for $m=0$,
which enables us to set $m=0$ without loss of generality.
From now on, we abbreviate the subscripts $\ell$ and $m$.
Explicit expressions for the background-dependent coefficients are given in the appendix \ref{app-a}.
We first notice that, due to the parity violating nature of the CS term, 
there appear mixing terms of odd and even perturbations in the action.
For the mixing terms that do not contain $\delta W$,
all the coefficients vanish if $W(r)$ does not depend on $r$.
This reflects the fact that the Chern-Simons term $C$ is a total derivative and affects the field equations of motion only when $W$, which multiplies $C$, depends on $r$.

This action shows that not all of the variables are dynamical.
Actually, we see that $H_0,~H_1$ and $\delta W$ are auxiliary fields.
Therefore, they can be eliminated from the action by using their equations of motion.
Since $H_0$ appears only linearly, the variation with respect to it gives a constraint among the other fields.
We eliminate $H_2$ by using this constraint.
The variation with respect to $H_1$ and $\delta W$ gives the equation of motion of each variable, respectively.
After substituting the constraints and many integration by parts, we end up
with the following Lagrangian density:
\begin{eqnarray}
{\cal L}&=&p_1 {\ddot h_1}^2+p_2 {\ddot h_1} (r {\dot h_0}'-2{\dot h_0})+p_3 {\dot h_0}'^2+p_4 {\dot h_0}^2+p_5 {\dot h_1}^2+p_6 {\dot {\delta F}}^2+p_7 {\dot \beta}^2+p_8 {\dot h_0} {\dot {\delta F}}+p_9 {\dot h_0}{\dot \beta}+p_{10} {\dot \beta}{\dot {\delta F}}+p_{11} h_0'^2 \nonumber \\
&&+p_{12} {\delta F}'^2 +p_{13} \beta'^2+p_{14} h_0' \delta F'+p_{15} h_0' \beta'+p_{16} \beta' \delta F'+p_{17} h_0' {\dot h_1}+p_{18} {\dot h_0}h_1+p_{19} h_0' \delta F+p_{20} h_0' \beta \nonumber \\
&&+p_{21} {\dot h_1}\delta F+p_{22} {\dot h_1}\beta+p_{23} \delta F \beta'+p_{24} h_0^2+h_0(p_{25} \delta F+p_{26} \beta )+p_{27} h_1^2+p_{28} \delta F^2+p_{29} \delta F \beta+p_{30} \beta^2. \label{fin-lag}
\end{eqnarray}
Since all the fields $(h_0,~h_1,~\beta,~\delta F)$ have time derivatives that are not removed by any
integration by parts, all of them are dynamical fields.
Hence, this is our final Lagrangian.
Explicit expressions of the background dependent coefficients $p_1,\cdots$ are given in the appendix \ref{app-p}.
However, since most of $p_1,\cdots$ have very long expressions,
we only provide them in a form truncated at the leading order in $F',~W',~F_\lambda,~F_s$ and $W_s$
all of which vanish in the GR limit.

In addition to the mixing terms of odd and even modes, 
there is another big qualitative difference between the above Lagrangian and that in $f(R)$ gravity.
The above Lagrangian contains a term ${\ddot h_1}^2$. 
This term results in fourth order differential equations for $h_1$ with respect to time,
\begin{equation}
2p_1 \frac{\partial^4 h_1}{\partial t^4}+p_2 \left( r \frac{\partial^3 h_1'}{\partial t^3}-2 \frac{\partial^3 h_1}{\partial t^3} \right)-2 p_5 {\ddot h_1}-p_{17} {\dot h_0'}+p_{18} {\dot h_0}-p_{21} {\dot {\delta F}}-p_{22} {\dot \beta}+2 p_{27} h_1=0.
\end{equation}
In $f(R)$ gravity, there are no second derivative terms in the Lagrangian and the resulting equations of motion are second order.
We can confirm this fact by looking at the explicit expressions for $p_1,~p_2$ and $p_3$, which are given by
\begin{equation}
p_1=-\frac{32 \pi  \ell (\ell+1) M_P^2 W'^2}{(2 \ell+1) F \left(\frac{A}{B}\right)^{3/2}},~~~p_2=-\frac{2p_1}{r},~~~p_3=p_1. \label{p1-p3}
\end{equation}
For $f(R)$ gravity, we have $W'=0$.
Therefore, $p_1,~p_2$ and $p_3$ vanish and the Lagrangian does not contain second derivative terms.

The presence of the ${\ddot h_1}$ term in the general $f(R,C)$ gravity is a signal that the theory is plagued by instability \cite{Woodard:2006nt}.
Actually, we can show that Hamiltonian corresponding to Eq.~(\ref{fin-lag}) is not bounded from below.
To construct the Hamiltonian, we find it useful to introduce a new field $q$ and rewrite Eq.~(\ref{fin-lag}) as
\begin{eqnarray}
{\cal L}&=&-p_1 \bigg[ q^2+2 {\dot q} \left( {\dot h_1}-h_0'+\frac{2}{r}h_0 \right) \bigg]+\bigg[ p_4-2 \left( \frac{p_1}{r} \right)'-\frac{4p_1}{r^2} \bigg] {\dot h_0}^2+p_5 {\dot h_1}^2+p_6 {\dot {\delta F}}^2+p_7 {\dot \beta}^2+p_8 {\dot h_0} {\dot {\delta F}}+p_9 {\dot h_0}{\dot \beta} \nonumber \\
&&+p_{10} {\dot \beta}{\dot {\delta F}}+p_{11} h_0'^2 +p_{12} {\delta F}'^2 +p_{13} \beta'^2+p_{14} h_0' \delta F'+p_{15} h_0' \beta'+p_{16} \beta' \delta F'+p_{17} h_0' {\dot h_1}+p_{18} {\dot h_0}h_1+p_{19} h_0' \delta F \nonumber \\
&&+p_{20} h_0' \beta + {\dot h_1}( p_{21}\delta F+p_{22}\beta )+p_{23} \delta F \beta'+p_{24} h_0^2+h_0(p_{25} \delta F+p_{26} \beta )+p_{27} h_1^2+p_{28} \delta F^2+p_{29} \delta F \beta+p_{30} \beta^2, \label{fin-lag2}
\end{eqnarray}
where we have used relations (\ref{p1-p3}).
It can be confirmed that Eq.~(\ref{fin-lag2}) reduces to Eq.~(\ref{fin-lag}) after eliminating $q$ by using its equation of motion.
The new Lagrangian (\ref{fin-lag2}) is much more familiar than Eq.~(\ref{fin-lag}) since it does not contain either the second time derivative term nor the mixing derivative term ${\dot h_0}'$.
Now, we can use the standard canonical formalism to construct the Hamiltonian.
The conjugate momenta are defined by
\begin{eqnarray}
&&\pi_q = \frac{\partial {\cal L}}{\partial {\dot q}}=-2p_1 \left( {\dot h_1}-h_0'+\frac{2}{r} h_0 \right), \\
&&\pi_0= \frac{\partial {\cal L}}{\partial {\dot h_0}}=2 \bigg[ p_4-2 \left( \frac{p_1}{r} \right)'-\frac{4p_1}{r^2} \bigg] {\dot h_0}+p_8 {\dot {\delta F}}+p_9 {\dot \beta}+p_{18} h_1, \\
&&\pi_1 =\frac{\partial {\cal L}}{\partial {\dot h_1}}=-2p_1 {\dot q}+2p_5 {\dot h_1}+p_{17} h_0'+p_{21} \delta F+p_{22} \beta, \\
&&\pi_F=\frac{\partial {\cal L}}{\partial {\dot {\delta F}}}=2p_6 {\dot {\delta F}}+p_8 {\dot h_0}+p_{10} {\dot \beta}, \\
&&\pi_\beta=\frac{\partial {\cal L}}{\partial {\dot \beta}}=2p_7 {\dot \beta}+p_9 {\dot h_0}+p_{10} {\dot {\delta F}}.
\end{eqnarray}
We can always solve these equations in terms of the field time derivatives since a Jacobian of these transformations is given by
\begin{equation}
\det \left(  \frac{\partial^2 \pi_i}{\partial {\dot q_i} \partial {\dot q_j}} \right) =\frac{805306368 \pi ^5 \ell^4 (\ell+1)^4 \left(\ell^2+\ell-2\right)^2 M_P^{10} B^5 \sqrt{\frac{A}{B}} W'^6}{(2 \ell+1)^5 A^4 F \left(r B A' \left(r F'+2 F\right)+2 A \left(F \left(-2 B+\ell^2+\ell \right)-r B F'\right)\right)^2},
\end{equation}
which is not zero in general (the special case where $W$ is a constant will be considered later).
Here we defined $q_1=q,~q_2=h_0,~q_3=h_1,~q_4=\delta F,~q_5=\beta$.
Therefore, the Lagrangian is not singular and there are no primary constraints among the canonical variables.
The Hamiltonian is then given by
\begin{eqnarray}
H=\int dr~{\cal H}=\int dr~ \left( \pi_q {\dot q}+\pi_0 {\dot h_0}+ \pi_1 {\dot h_1}+\pi_F {\dot {\delta F}}+\pi_\beta {\dot \beta}-{\cal L}  \right).
\end{eqnarray}
It can be confirmed that a matrix $K_{ij}$, defined by the momentum part of the Hamiltonian, 
\begin{equation}
{\cal H} \supset K_{ij} \pi_i \pi_j+\cdots,
\end{equation}
where $\cdots$ represents terms that are not quadratic in $\pi_i$, has a vanishing component for $K_{33}$ and a nonvanishing component for $K_{13}$,
\begin{equation}
K_{33}=0,~~~~~K_{13}=-\frac{1}{4p_1}.
\end{equation}
In addition, we find $\{ \pi_1,~\pi_3 \}$ and $\{ \pi_2,~\pi_4,~\pi_5 \}$ are decoupled from each other.
Combined with these facts, the subspace of the Hamiltonian spanned by $\pi_1$ and $\pi_3$ yields a negative determinant of the corresponding subkinetic matrix,
\begin{equation}
K_{11} K_{33}-K_{13}^2=-K_{13}^2<0.
\end{equation}
This means that the Hamiltonian can take arbitrary negative values by suitably choosing the values of $\pi_1$ and $\pi_3$ and hence it is not bounded from below.
This result shows that the general $f(R,C)$ gravity has the problem of having a ghost around the static and spherically symmetric background and provides a severe condition on the functional form of $f(R,C)$.

There are two possible cases where the presence of the ghost does not become problematic.
The first one is to assume that the $f(R,C)$ theory under consideration is an effective theory which is valid only on length scales larger than a certain length $d_c$.
From this point of view, the presence of the ghost does not matter if its mass is larger than the energy scale $d_c^{-1}$ since the dynamics of the ghost cannot be described by the low energy $f(R,C)$ theory.
The more fundamental theory which is valid above $d_c^{-1}$ may cure the problem.
The mass of the ghost can be evaluated as follows.
Neglecting the gradient terms for the fields which are not important here, the Hamiltonian density can be written as
\begin{equation}
{\cal H}=K_{ij} \pi_i \pi_j+B_{ij} \pi_i q_j+M_{ij} q_i q_j.
\end{equation}
There are mixing terms between $\pi_i$ and $q_j$ that are represented by a matrix $B_{ij}$.
Let us first eliminate the mixing terms by the following canonical transformation:
\begin{eqnarray}
&&\pi_i =P_i -\frac{1}{2} {\left( K^{-1} B \right)}_{ij} Q_j, \\
&&q_i=Q_i.
\end{eqnarray}
In terms of the new canonical variables, the Hamiltonian density can be written as
\begin{equation}
{\cal H}=K_{ij}P_i P_j+{\left( M-\frac{1}{4}B^T K^{-1}B \right)}_{ij} Q_i Q_j.
\end{equation}
It turns out that the new mass matrix does not mix $\{ Q_1,~Q_3 \}$ and $\{ Q_2,~Q_4,~Q_5 \}$.
Therefore, the first and the third canonical fields form a closed system.
The corresponding sub-Hamiltonian is given by
\begin{equation}
{\cal H}_{\rm sub}=K_{11}P_1^2+2K_{13} P_1 P_3+{\bar M}_{11} Q_1^2+{\bar M}_{33} Q_3^2,
\end{equation}
where ${\bar M} \equiv M-\frac{1}{4}B^T K^{-1}B$ and each element is given by
\begin{eqnarray}
&&K_{11}=\frac{(2 \ell+1) \left(64 \ell (\ell+1) F'^2  F_s^2-r^4 F_\lambda \right)}{2048 \pi \ell (\ell+1) M_P^2 r^4 F_\lambda W'^4},~~~K_{13}=\frac{2 \ell+1}{128 \pi \ell \left(\ell+1 \right) M_P^2 W'^2}, \nonumber \\
&&{\bar M}_{11}=-\frac{32 \pi  \ell (\ell+1) M_P^2 W'^2}{2 \ell+1},~~~{\bar M}_{33}=-\frac{3 \pi \ell (\ell+1) M_P^2 \left(r^4 F_\lambda -64 \ell (\ell+1) F_s^2 F'^2\right)^2}{32 \left(2 \ell^3+3 \ell^2-3 \ell-2\right) r^8 F_\lambda^2 W'^2}.
\end{eqnarray}
We notice that both ${\bar M}_{11}$ and ${\bar M}_{33}$ are negative definite, which means both $Q_1$ and $Q_3$ become tachyonic as well if the magnitudes of their masses are smaller than $d_c^{-1}$.
We can make the sub-Hamiltonian, which is a sum of two independent harmonic oscillators, by the following canonical transformation,
\begin{eqnarray}
&&P_1=\sqrt{-{\bar M}_{11}} \left( {\bar P}_1 \cos \delta + {\bar P}_3 \sin \delta  \right),~~~P_3=\sqrt{-{\bar M}_{33}} \left( -{\bar P}_1 \sin \delta + {\bar P}_3 \cos \delta \right), \nonumber \\
&&Q_1=\frac{1}{\sqrt{-{\bar M}_{11}}} \left( {\bar Q}_1 \cos \delta + {\bar Q}_3 \sin \delta \right),~~~Q_3=\frac{1}{\sqrt{-{\bar M}_{33}}} \left( - {\bar Q}_1 \sin \delta + {\bar Q}_3 \cos \delta \right),
\end{eqnarray}
where $\delta$ is determined from the equation
\begin{equation}
\tan 2\delta = -\frac{2 K_{13}}{K_{11}} \sqrt{\frac{{\bar M}_{33}}{{\bar M}_{11}}}=\sqrt{\frac{3}{\ell^2+\ell-2}}~ {\rm sgn} \left( r^4 F_\lambda -64 \ell (\ell+1) F_s^2 F'^2 \right).
\end{equation}
Here, the sign function is defined as ${\rm sgn}(x)= +1,0,-1$ for $x>0,x=0,x<0$, respectively.
The new sub-Hamiltonian is then given by
\begin{equation}
{\cal H}_{\rm sub}=-\cos \delta \left( K_{11} {\bar M}_{11} \cos \delta+2K_{13} \sqrt{{\bar M}_{11}{\bar M}_{33}} \sin \delta \right) {\bar P}_1^2-\sin \delta \left( K_{11} {\bar M}_{11} \sin \delta-2K_{13} \sqrt{{\bar M}_{11}{\bar M}_{33}} \cos \delta \right) {\bar P}_3^2-{\bar Q}_1^2-{\bar Q}_3^2. \nonumber
\end{equation}
The coefficients in front of ${\bar P}_1$ and ${\bar P}_3$ must be larger than $d_c^{-2}$ in order for those fields to be in the high energy regime where the effective $f(R,C)$ theory does not work.
This leads to a condition,
\begin{equation}
| K_{11} M_{11} | \simeq \bigg| \frac{r^4 F_\lambda -64 \ell (\ell+1) F_s^2 F'^2}{64 r^4 F_\lambda W'^2}\bigg| \gtrsim d_c^{-2}.
\end{equation}
In particular, when the second term in the numerator is negligible, we obtain the very simple condition for $W'$,
\begin{equation}
|W'| \lesssim d_c. \label{condition-W}
\end{equation}
Since $W'$, which has dimensions of length, represents how large the effects of the Chern-Simons term are, this condition says these effects are suppressed on distances larger than $d_c$.

The second possibility where the presence of the ghost does not become problematic is that $f(R,C)$ belongs to the special class in which $W'=0$ is satisfied identically.
Using the background metric, $W'$ can be written as
\begin{equation}
W'=F_s R'+W_s C'=F_s R',
\end{equation}
where we have used an identity $C=0$ for the background metric.
Therefore, if $f(R,C)$ satisfies either $F_s=0$ or $R={\rm const}$, we have $W'=0$ identically.
For example, $F_s=0$ is trivially satisfied if $f(R,C)$ takes a separable form, {\it i.e.}, $f(R,C)=f_1(R)+f_2 (C)$, where $f_1$ and $f_2$ are arbitrary functions of $R$ and $C$, respectively.
$f(R)$ gravity is included in this case.
The second case $R={\rm const}$ is satisfied, for example, if the Schwarzschild metric is a solution of the model.
In either case, we have still many $f(R,C)$ theories.
We deal with this class of $f(R,C)$ theories in the next section.

\section{Study of special cases with $W'=0$.}
As we have shown in the previous section, the general $f(R,C)$ theories with nonvanishing $W'$ have the problem of instability. 
Thus, the cases with $W'=0$ are more phenomenologically interesting and deserve further investigation.
In this section, we study the second order perturbation again for this case.
We clarify the number of propagating modes and derive dispersion relations for them.
First, we analyze the general case for $\ell \ge 2$, and consider the special case for $\ell = 0$ and $\ell = 1$.

\label{sec-sp}
\subsection{Second order action again}
The Lagrangian for the general $f(R,C)$ theories (\ref{fin-lag}) can be also used for the special case $W'=0$.
Since the condition $W'=0$ makes some terms identically vanish and the resulting Lagrangian is greatly simplified, it is better to write the simplified Lagrangian,
\begin{eqnarray}
{\cal L}&=&q_1 {\left( h_0'-{\dot h_1} \right)}^2+q_2 {\dot {\delta F}}^2+q_3 {\delta F'}^2+q_4 {\dot {\delta F}} {\dot \beta}+q_5 \delta F' \beta'+q_6 {\dot \beta}^2+q_7 \beta'^2+q_8 h_0{\dot h_1} \nonumber \\
&&+q_9 \delta F (h_0'-{\dot h_1}) +q_{10} \beta \delta F'+q_{11} h_0^2+q_{12} h_0 \delta F+q_{13} h_1^2+q_{14} \delta F^2+q_{15} \beta \delta F+q_{16} \beta^2, \label{sp-lag}
\end{eqnarray}
where the background-dependent coefficient can be read from $p_1,\cdots$ by imposing a condition $W'=0$.
The odd and even modes are still coupled.
The background-dependent coefficients responsible for the coupling are given by
\begin{equation}
q_9=\frac{16 \pi \ell (\ell+1) M_P^2 F_s \left(r B A' \left(r F'+2 F \right)-2 A \left(r B F'+2 (B-1) F\right)\right)}{(2 \ell+1) r^2 A F F_\lambda},~~~q_{12}=-\frac{2}{r} q_9.
\end{equation}
Interestingly, the odd and even modes are decoupled if $F_s=0$. 
Furthermore, if $F_s=0$, we verified that all the coefficients $q_i$ for the even modes are exactly the same as those for the $f(R)$ theories.
On the other hand, the coefficients $q_i$ for the odd modes depend on $W_s$.
Therefore, for the $f(R,C)$ theories having the property $F_s=0$, the signature of the parity violation appears only in the odd modes. 
In the following analysis, we do not assume $F_s=0$ to keep the procedure as general as possible.

Similar to what we did in the previous section, we can introduce a new variable $q$ to write the above Lagrangian as
\begin{eqnarray}
{\cal L}&=&q_1 \bigg[ 2q \left( h_0'-{\dot h_1}+\frac{2}{r} h_0 \right)-q^2 \bigg]+q_2 {\dot {\delta F}}^2+q_3 {\delta F'}^2+q_4 {\dot {\delta F}} {\dot \beta}+q_5 \delta F' \beta'+q_6 {\dot \beta}^2+q_7 \beta'^2 \nonumber \\
&&+q_9 \delta F (h_0'-{\dot h_1}) +q_{10} \beta \delta F'+\left( q_{11}-\frac{1}{2} q_8'-\frac{q_8}{r} \right) h_0^2+q_{12} h_0 \delta F+q_{13} h_1^2+q_{14} \delta F^2+q_{15} \beta \delta F+q_{16} \beta^2. \label{sp-lag2}
\end{eqnarray}
Here we have used a relation $q_8=2q_1$.
We can verify that the new Lagrangian reduces to the original one by eliminating $q$ using its equation of motion.
The point of the new Lagrangian is that it contains derivatives of $h_0$ and $h_1$ up to at most first order while the original one contains $h'_0{}^2$ and ${\dot h_1}^2$.
Therefore, by doing integration by parts, we can rewrite the new Lagrangian in such a way that any derivative of $h_0$ and $h_1$ does not appear anymore.
After this procedure, both $h_0$ and $h_1$ become auxiliary variables.
For $\ell \ge 2$, quadratic terms in $h_0$ and $h_1$ exist and we can eliminate $h_0$ and $h_1$ by using their equations of motion.
After this, we finally obtain a Lagrangian which contains only $q,~\delta F$ and $\beta$.
We find that the final Lagrangian can be formally written as
\begin{equation}
{\cal L}=k_{ij} {\dot q_i} {\dot q_j}-d_{ij} q_i' q_j'-e_{ij} q_i' q_j-m_{ij} q_i q_j, \label{sp-lag3}
\end{equation}
where we have defined $(q_1,q_2,q_3)=(\delta F,\beta,q)$.
The cases where $\ell$ is either $0$ or $1$ will be studied later.

A determinant of the kinetic matrix $k_{ij}$ is found to be
\begin{equation}
\det (k_{ij})=\frac{384 \pi ^3 \ell^2 (\ell+1)^2 M_P^6 r^4 Y^2}{(2 \ell+1)^3 A^3 F^3 \left(\frac{A}{B}\right)^{3/2} \left(r B A' \left(r F'+2 F\right)+2 A \left(F \left(-2 B+\ell^2+\ell\right)-r B F'\right)\right)^2}, \label{sp-det}
\end{equation}
where $Y$ is defined by
\begin{equation}
Y=\frac{16 \ell (\ell+1) \left( F_\lambda W_s-F_s^2\right)
   \left(r B A' \left(r F'+2 F\right)-2 A \left(r B F'+2 (B-1) F\right)\right)^2}{r^6 F_\lambda }+A^2 F^3.
\end{equation}
This is not zero in general.
Therefore, all the variables are dynamical and there are three propagating modes, one of which is odd ({\it i.e.}, $q$) and the remaining two are even ({\it i.e.}, $\delta F$ and $\beta$).
This structure is the same as that of the $f(R)$ gravity theories where there is one propagating odd mode and two propagating even modes.
This result shows that the condition $W'=0$ kills all the pathological modes which, as we found in the previous section, exist in the general $f(R,C)$ theories.
To see if the modes are ghosts or not, we find it convenient to evaluate $k_{33}$ and $k_{22} k_{33}-k_{23}^2$,
\begin{eqnarray}
k_{33}&=&\frac{2 \pi  \ell (\ell+1)^2 M_P^2 r^2 Y^2}{\left(2 \ell^4+5 \ell^3-5 \ell-2\right) A^5 F^5 \sqrt{\frac{A}{B}}}, \\
k_{22} k_{33}-k_{23}^2&=&\frac{32 \pi ^2 \ell^2 (\ell+1)^3 M_P^4 r^2 Y^2 B^2 \left(\left(2 \ell^2+2 \ell-1\right) r^2 F'^2+2 \left(\ell^2+\ell-2\right) r F
   F'+2 \left(\ell^2+\ell-2\right) F^2\right)}{(2 \ell+1)^2 \left( \ell^3+2 \ell^2-\ell-2\right) A^4 F^4 \left(r B A' \left(r F'+2
   F\right)+2 A \left(F \left(-2 B+\ell^2+\ell \right)-r B F'\right)\right)^2}.
\end{eqnarray}
To avoid the ghost, we need to impose $F>0$.
\footnote{There is also another reason the condition $F>0$ is required.
As we can see from Eq.~(\ref{act2}), $G_N F^{-1}$ gives the effective gravitational constant. 
Therefore, negative $F$ yields repulsive gravitational force, which clearly contradicts our experiences.}
With this condition, both $\det (k_{ij})$ and $k_{33}$ are clearly positive definite.
After a short calculation, it can be verified that $k_{22} k_{33}-k_{23}^2$ is also positive definite.
Therefore, as is the case with $f(R)$ theories, $F>0$ is the no-ghost condition for $f(R,C)$
theories that satisfy $W'=0$.

We can derive the dispersion relations for the three modes from an equation,
\begin{equation}
\det (-\omega^2 k_{ij}+k^2 d_{ij})=0.
\end{equation}
Explicit calculation gives 
\begin{equation}
\det (-\omega^2 k_{ij}+k^2 d_{ij})=\frac{768 \pi^3 \ell^2 \left(-\ell^3-2 \ell^2+\ell+2\right)^2 M_P^6 r B F^3 q_1^2 \sqrt{\frac{A}{B}} \left(\omega
   ^2-k^2 A B\right)^3}{(2 \ell+1)^3 q_{13} \left(2 r q_{11}-r q_8'-2 q_8 \right) \left(r B
   A' \left(r F'+2 F\right)+2 A \left(F \left(-2 B+\ell^2+\ell \right)-r B F'\right)\right)^2}.
\end{equation}
We see all the modes obey a dispersion relation $\omega^2=AB k^2$.
The appearance of the factor $AB$ is due to the fact that $t$ and $r$ are coordinate time and distance.
In terms of the physical time and distance, the dispersion relation says all
the modes propagate at the speed of light, which is exactly the same as in the case of $f(R)$ theories.
Although there are no new contributions to the propagation speeds due to the Chern-Simons term, inclusion of the Chern-Simons yields a new correction to the propagation properties even when $W'=0$.
We find that the off-diagonal term $k_{13}$ is nonvanishing,
\begin{equation}
k_{13}=\frac{8 \pi \ell (\ell+1)^2 M_P^2 Y F_s \left(r B A' \left(r F'+2 F\right)-2 A \left(r B F'+2 (B-1) F\right)\right)}{(2 \ell+1) \left( \ell^3+2 \ell^2-\ell-2\right) A^4 F^4 F_\lambda}.
\end{equation}
We also find that $d_{13},~e_{31}$ and $m_{13}$ are nonvanishing in general.
This means that $\delta F$ and $q$ are coupled.
On the other hand, there is no direct coupling between $\beta$ and $q$.
The coupling between $\delta F$ and $q$ means that we cannot consider the propagation of the odd and even modes separately as we can do in the case of $f(R)$ theories, which is a clear difference from $f(R)$ theories.
This shows the potential usefulness of using the distinct nature of mode propagation in $f(R,C)$ theories for putting constraints on $f(R,C)$ models using observations of gravitational waves from compact astrophysical objects, when these become available in the future.

We can also evaluate the mass for each eigenmode.
However, since each matrix element of $m_{ij}$ is too lengthy to obtain
analytic expressions for the mass eigenvalues, we will make an assumption that the background is very close to GR, {\it i.e.}, $F=1$, $A=B=1-\frac{r_g}{r}$ and also expand the eigenvalues in $\varepsilon \equiv \frac{r_g}{r}$ (weak field approximation).  
Under these assumptions, three eigenvalues are given by
\begin{eqnarray}
&&m_1^2=\frac{1}{3F_\lambda}-\frac{\ell^2 (\ell+1)^2 r_g-2 \left( \ell^2+\ell-2\right)^2 \left( \ell^2+\ell+2\right) r}{3 \left( \ell^2+\ell-2\right)^2 r^3}-\frac{64 \ell \left( \ell^5+3 \ell^4+7 \ell^3+9 \ell^2-4\right) r_g^2 F_s^2}{\left( \ell^2+\ell-2\right)^2 r^8 F_\lambda^2}+{\cal O} \left( \varepsilon^3 \right), \\
&&m_2^2=\frac{\ell (\ell+1)}{r^2}-\frac{\ell^2 (\ell+1)^2 r_g}{\left( \ell^2+\ell-2\right)^2 r^3}+{\cal O}(\varepsilon^3), \\
&&m_3^2=\frac{\ell^2+\ell+4}{r^2}-\frac{192 \ell \left( \ell^5+3 \ell^4+7 \ell^3+9 \ell^2-4\right) r_g^2 F_s^2}{\left( \ell^2+\ell-2\right)^2 r^8 F_\lambda^2}+{\cal O}(\varepsilon^3).
\end{eqnarray}
Since $m_1^2$ is inversely proportional to $F_\lambda$ at leading order, this mode corresponds to the scalar graviton that exists in the general $f(R)$ theories.
To avoid the tachyonic mode, we need to impose a condition $F_\lambda>0$.
The Chern-Simons corrections, $F_s$, appear in $m_1^2$ and $m_3^2$, but only in a combination with $r_g$.
This means those corrections are important only in the vicinity of the BH and are suppressed compared to the standard terms that exist in GR far from the BH.

As we mentioned earlier, the results we derived do not apply to $\ell=0$ and $\ell=1$ modes since
we cannot solve $h_0$ and $h_1$ to go from Eq.~(\ref{sp-lag2}) to Eq.~(\ref{sp-lag3}).
In the following, we consider the $\ell=0$ and $\ell =1$ cases, separately.

\subsection{Monopole perturbation:~$\ell =0$}
The procedure we took from Eq.~(\ref{bare-lag}) to Eq.~(\ref{fin-lag}) cannot apply for the monopole case since $b_1$ identically vanishes for $\ell=0$.
Therefore, let us start again from Eq.~(\ref{bare-lag}).
For the monopole case, the odd modes and $\alpha$ identically vanish,
which drastically simplifies the Lagrangian.
In addition to this, the following relations among the background-dependent coefficients hold for $\ell=0$,
\begin{equation}
a_5=\left( \frac{a_7 b_3}{b_2} \right)',~~~a_6=\frac{a_7 b_3}{b_2}+a_7,~~~\frac{a_7 b_4}{b_2}=a_2.
\end{equation}
Using these relations, we find that Eq.~(\ref{bare-lag}) reduces to
\begin{eqnarray}
{\cal L}&=&H_0 \left( a_2 H_2+\frac{a_7 b_3}{b_2} \delta F+a_7 \delta F' \right)'+\frac{b_2}{a_7} H_1 {\left( a_2 H_2+\frac{a_7 b_3}{b_2} \delta F+a_7 \delta F' \right)}^\cdot+c_1 H_2^2+c_3 H_2 \delta F'+c_4 H_2 \delta F \nonumber \\
&&+c_8 {\dot H_2} {\dot {\delta F}}+f_1 \delta F^2+f_2 \delta F \delta W+f_3 \delta W^2. \label{mono-lag}
\end{eqnarray}
We can again eliminate $\delta W$ by using its equation of motion and the remaining fields are $(H_0,H_1,H_2,\delta F)$.
For $H_0$ and $H_1$, they appear only linearly.
Therefore, they are Lagrange multipliers and their equations of motion yield constraints among the other fields.
At first glance, we derived two constraints seem to over-constrain $H_2$ and $\delta F$.
However, the two constraints are not independent of each other and we get only a single constraint which is given by
\begin{equation}
a_2 H_2+\frac{a_7 b_3}{b_2} \delta F+a_7 \delta F'=-C_1, \label{mono-constraint}
\end{equation}
where $C_1$ is an integration constant.
This gives $H_2$ in terms of $\delta F$.

It is important to notice here that the gauge conditions $\beta=0,~K=0$ and $G=0$ we have used to fix the even modes do not work for the $\ell=0$ case since $h_{ta}$, $h_{ra}$ and the second term in Eq.~(\ref{hab}) identically vanish.
A condition that remains meaningful is $K=0$, by which $R_{\ell m}$ is completely fixed and we still have another gauge transformation, $T_{\ell m} \equiv T_0$.
We can always set $H_1=0$ by properly choosing $T_0$. 
However, this condition still allows a remaining gauge degree of freedom of the form $T_0= C_2(t) A(r)$, where $C_2(t)$ is an arbitrary function.
Correspondingly, $H_0$ contains a gauge mode which is given by $2\dot C_2(t)$.
Variation of Eq.~(\ref{mono-lag}) with respect to $H_2$ gives a following equation:
\begin{equation}
a_2 H_0'=2 c_1 H_2+c_3 \delta F'+c_4 \delta F. \label{H0-constraint}
\end{equation}
Since $H_2$ is written in terms of $\delta F$ by Eq.~(\ref{mono-constraint}), $H_0$ is also determined by Eq.~(\ref{H0-constraint}) once $\delta F$ is known.
A homogeneous solution of $a_2 H_0'=0$ is $H_0=C_3(t)$, where $C_3(t)$ is an arbitrary function, is a gauge mode and hence can be set to zero.

Putting this constraint (\ref{mono-constraint}) into Eq.~(\ref{mono-lag}) yields a Lagrangian which consists only of $\delta F$,
\begin{equation}
{\cal L}=v_1 \left( {\dot {\delta F}}^2-AB {\delta F'}^2 \right)+v_2 \delta F^2+C_1 v_3 \delta F,
\end{equation}
where $v_1,~v_2$ and $v_3$ are given by
\begin{eqnarray}
v_1&=&\frac{24 \pi M_P^2 r^2 F}{\sqrt{AB}\left(r F'+2 F\right)^2}, \\
v_2&=&-\frac{2 \pi M_P^2 r^2 \sqrt{A}}{\sqrt{B} F_\lambda}-\frac{\pi M_P^2 \left(r A A' \left(9 r B'+4 B\right)-9 r^2 B A'^2+2 A^2 \left(3 r^2 B''+6 r B'+2 B-2\right)\right)}{2 B^2 F \left(\frac{A}{B}\right)^{3/2}} + \cdots, \\
v_3&=&-\frac{2 \pi M_P^2 \left( 4 A^2 \left(3 r F B' \left(r F'+2 F\right)+2 B \left(F-r
   F'\right)^2\right) -r B A' \left(r F'+2 F\right) \left( rA' (rF'+2F)+2A(4rF'+5F) ) \right) \right)}{A^{3/2} \sqrt{B} \left(r F'+2 F\right)^2}
\end{eqnarray}
Since $v_2$ has a very long expression, we have kept only terms which either diverge or remain finite in the GR limit $F_\lambda \to 0$.
The last term which is linear in $\delta F$ acts as a source for $\delta F$.
It is clear from the final Lagrangian that there is only one propagating mode which does not exist in GR.
This mode appears when $F$ is allowed to fluctuate, as in the case of $f(R)$ theories.
The propagation speed is the velocity of light.
The no-ghost condition $v_1>0$ is satisfied if $F>0$.
Therefore, as is the case with $\ell \ge 2$, $F>0$ is the no-ghost condition.
In order not to make the mode tachyonic, we need to impose another condition $F_\lambda >0$.

Let us finally comment on the physical meaning of the other mode, which is accompanied by the constant $C_1$ in Eq.~(\ref{mono-constraint}).
This can be understood by considering the case of GR.
In this case, we have $\delta F=0$ and $a_2=-4M_P^2 \pi rA$ and $c_1=2M_P^2 \pi$.
Then, Eq.~(\ref{mono-constraint}) becomes
\begin{equation}
H_2 = \frac{C_1}{4M_P^2 \pi rA}.
\end{equation}
Putting this into Eq.~(\ref{H0-constraint}) gives
\begin{equation}
H_0 = \frac{C_1}{4M_P^2 \pi rA}=H_2.
\end{equation}
These solutions correspond to a metric,
\begin{equation}
ds^2=-\left( 1-\frac{2G_N (M+C_1)}{r} \right) dt^2+{\left( 1-\frac{2G_N (M+C_1)}{r} \right)}^{-1} dr^2+r^2 (d\theta^2+\sin^2 \theta~d\varphi^2).
\end{equation}
Therefore, $C_1$ just shifts a parameter of the background solution into another one. 
Physically, $C_1$ represents shift of the BH mass.

\subsection{Dipole perturbation:~$\ell =1$}
Let us start from Eq.~(\ref{sp-lag}).
For the dipole case, we find that the relation $q_2 q_6-\frac{1}{4} q_4^2=0$ holds.
This means only a linear combination of $\delta F$ and $\beta$ can be a dynamical variable.
To construct a dynamical variable, let us do a field transformation from $\delta F$ to $\psi$ by
\begin{equation}
\psi = \delta F+\frac{q_4}{2q_2} \beta.
\end{equation}
When written in terms of the new variables, no ${\dot \beta}$ appears in the Lagrangian.
At the same time, we find all the $\beta'$ terms go away. 
This is due to another relation, $q_3 q_7-\frac{1}{4} q_5^2=0$.
Now, since $\beta$ is an auxiliary field, we can eliminate it by using its equation of motion.
After this, the Lagrangian consists of $\psi$, $h_0$ and $h_1$, which can be formally written as
\begin{equation}
{\cal L}=s_1 \left( {\dot \psi}^2-AB \psi'^2 \right)+s_2 \psi^2+\frac{2}{r} \left( s_3'+\frac{s_3}{r} \right) h_0^2+s_3 h_0'^2+s_4 {\dot h_1} h_0-\frac{r}{2} s_4 {\dot h_1} h_0'+s_5 h_1^2+s_6 {\dot h_1}^2.
\end{equation}
Interestingly, there is no coupling between the odd and even modes.

Let us first focus on the even mode $\psi$.
We find that $\psi$ propagates at the speed of light.
A coefficient in front of the kinetic term is given by
\begin{equation}
s_1=\frac{2\pi M_P^2r^2 \sqrt{\frac{A}{B}} \left(r BA' \left(rF'+2F\right)-4A(B-1)F\right)^2}{AF \left(r B A' \left(r F'+2 F\right)-2 A \left(r B F'+2 (B-1) F\right)\right)^2}.
\end{equation}
Therefore, as is the case with $\ell \ge 2$, $F>0$ is the no-ghost condition.
The mass term is very complicated in general, but if the background is very close to that in GR, we find
\begin{equation}
s_2=-\frac{M_P^2 \pi r^2}{2F_\lambda}.
\end{equation}
In this case, the mode is not tachyonic if $F_\lambda >0$.

Let us next consider the odd modes $h_0$ and $h_1$.
The equations of motion for them are 
\begin{eqnarray}
&&\left( 2s_3 h_0'-\frac{r}{2} s_4 {\dot h_1} \right)'=s_4 {\dot h_1}+\frac{4}{r} \left( s_3'+\frac{s_3}{r} \right) h_0, \label{eq-h0} \\
&&s_4 {\dot h_0}-\frac{r}{2} s_4 {\dot h_0}'+2s_6 {\ddot h_1}=2s_5 h_1. \label{eq-h1}
\end{eqnarray}
It is important to notice that the Regge-Wheeler gauge $h_2=0$ we have used to fix the gauge for the odd modes does not work for $\ell=1$ case.
This is simply because $h_{ab}$ for the odd modes identically vanishes for $\ell=1$ [see Eq.~(\ref{odd-ab})].
Therefore, there is still one gauge degree of freedom in the odd modes.
We found it convenient to set $h_1=0$.
However, as is clear from Eq.~(\ref{h1-gauge}), the gauge is not completely fixed yet and
we have a remaining gauge which is represented by $\Lambda =C_1(t) r^2$, where $C_1(t)$ is an arbitrary function.

Under the gauge condition $h_1=0$, Eq.~(\ref{eq-h0}) and Eq.~(\ref{eq-h1}) become
\begin{eqnarray}
&&\left( s_3 h_0' \right)'=\frac{2}{r} \left( s_3'+\frac{s_3}{r} \right) h_0, \label{neweq-h0} \\
&&{\dot h_0}-\frac{r}{2} {\dot h_0}'=0. \label{neweq-h1}
\end{eqnarray}
The general solutions of Eq.~(\ref{neweq-h1}) are given by
\begin{equation}
h_0=C_2 (t) r^2+y(r), \label{sol-h0}
\end{equation}
where $C_2 (t)$ and $y(r)$ are arbitrary functions.
Clearly, the first term $C_2 (t) r^2$ is also a solution of Eq.~(\ref{neweq-h0}).
But, this term is a gauge mode because it can be removed away by using the remaining gauge degree 
of freedom $\Lambda =C_1(t) r^2$.
On the other hand, the second term $y(r)$ represents a physical solution.
The form of $y(r)$ is determined by the condition that $y(r)$ must satisfy Eq.~(\ref{neweq-h0}) which can be rewritten as
\begin{equation}
y''+\frac{s_3'}{s_3}y'-\frac{2}{r} \left( \frac{1}{r}+\frac{s_3'}{s_3} \right) y=0. \label{eq-y}
\end{equation}
Since this is a second order differential equation, there are two independent solutions.
Of course, one solution is a gauge mode which is proportional to $r^2$ and can be absorbed into the first term of Eq.~(\ref{sol-h0}).
What we want is the other solution.
To obtain a formal expression of the desired solution $y$, let us construct a following quantity out of $y$ and $r^2$:
\begin{equation}
\Omega=s_3 (r^2 y'-2r y). \label{wro}
\end{equation}
By using Eq.~(\ref{eq-y}), we can show that $\Omega'=0$.
Setting $\Omega={\rm const}$ allows us to formally solve Eq.~(\ref{wro}) in terms of $y$,
\begin{equation}
y=D_1 r^2 \int_{r_0}^r \frac{d{\tilde r}}{s_3({\tilde r}) {\tilde r}^4}, \label{sol-y}
\end{equation}
where $r_0$ and $D_1$ are constants.
Although there are two free constants we can choose, the shift of $y$ due to shift of $r_0$ from one value to another only results in the change of the first term of Eq.~(\ref{sol-h0}), which can be gauged away. 
On the other hand, solutions with different values of $D_1$ are not connected by the gauge transformation and hence $D_1$ represents a physical quantity.
The explicit expression for $s_3$ is given in the appendix \ref{app-s}.

Since $y$ in Eq.~(\ref{sol-y}) depends only on $r$ and corresponds to the perturbation of the $t-\varphi$ metric component, this perturbation represents a stationary spacetime which slightly deviates from the static spacetime.
Physically, it represents a spacetime metric around a slowly rotating BH, where the angular momentum of the rotating BH is taken into account as the perturbation.
Therefore $D_1$ is related to the angular momentum of the BH.

For demonstration, let us check that Eq.~(\ref{sol-y}) in the case of GR actually gives 
a Kerr metric expanded to first order in the angular momentum $J$,
\begin{equation}
ds^2_{\rm Kerr} \simeq -\left( 1-\frac{r_g}{r} \right) dt^2+\frac{dr^2}{1-\frac{r_g}{r}}+r^2 (d\theta^2+\sin^2 \theta d\varphi^2)-\frac{2r_g J}{Mr} \sin^2 \theta~ dt d \varphi,
\label{kerr}
\end{equation}
where $M$ is a mass of the BH.
For the case of GR, we find $s_3=\frac{4\pi}{3} M_P^2$.
Then, Eq.~(\ref{sol-y}) becomes $y=-\frac{D_1}{4\pi M_P^2 r}$ which, in terms of $h_{t \varphi}$, can
be written as
\begin{equation}
h_{t \varphi}=\frac{D_1}{4\pi M_P^2 r}\sin^2 \theta.
\end{equation} 
This actually coincides with the last term of Eq.~(\ref{kerr}) for $D_1=-2J$.

\section{Conclusion}
We have studied linear perturbations around the static, spherically-symmetric spacetime for general $f(R,C)$ theories, where $C$ is the parity violating Chern-Simons term.
By explicitly constructing the second order action, we showed that one odd mode appears in the action as a quadratic in its second time derivative.
Irrespective of its sign, this results in an Hamiltonian that is not bounded from below.
Therefore, the static and spherically symmetric spacetime is unstable in general $f(R,C)$ theories.
This gives a strong limit on any phenomenological gravitational model which violates parity.

We also showed that either $R={\rm const}$ or $\frac{\partial^2 f}{\partial R \partial C}=0$ for the background metric is a necessary and sufficient condition
to avoid the instability mentioned above.
For such theories, the number of propagating modes for $\ell \ge 2$ is three, one from the odd and the other two from the even.
Unlike in the case of $f(R)$ theories, those modes are coupled, which can be used as a distinctive feature to test the parity violating theories from observations.
All the modes propagate at the speed of light.
The no-ghost condition is $\frac{\partial f}{\partial R}>0$ and the no-tachyon condition is $\frac{\partial^2 f}{\partial R^2}>0$, which are the same as in the case of $f(R)$ theories.
For the monopole perturbations ($\ell=0$), we showed that there is one propagating mode and one nonpropagating mode.
Since the propagating mode amounts to the fluctuation of $\frac{\partial f}{\partial R}$, it does not exist in GR, but exists in $f(R)$ gravity. 
It propagates with the velocity of light. 
On the other hand, the nonpropagating mode amounts to the shift of the mass of BH. 
This mode also exists in GR, as it should be.
For the dipole perturbations ($\ell=1$), we found that the odd and the even modes completely decouple.
The odd mode does not propagate and depends only on $r$.
Physically, it corresponds to a slowly-rotating BH solution whose metric is linearized in its angular momentum.
On the other hand, there is one even mode which does not exist in GR.
This mode also propagates at the speed of light.
The no-ghost condition and no-tachyon condition for $\ell=0$ and $\ell=1$ are the same as for $\ell \ge 2$.

\begin{acknowledgments}
We would like to thank M.~Lake for reading our manuscript and providing useful suggestions.
This work was supported in part by JSPS (H.M.) and JSPS Grant-in-Aid for Fellows No.~1008477 (T.S.).
T.S. thanks the Centre for Cosmology, Particle Physics, and Phenomenology
at Universit\'{e} Catholique de Louvain for its hospitality during the completion of this work.
\end{acknowledgments}

\newpage

\appendix
\section{Expressions of $a_1,~a_2,\cdots$.}
\label{app-a}
\begin{eqnarray}
&&a_1=-\frac{2 \pi  M_P^2 \left(r B A' \left(r F'+2 F\right)+A \left(4 r B F'+F \left(2 B+\ell^2+\ell\right)\right)\right)}{(2 \ell+1) \sqrt{A B}},~~~a_2=-\frac{2 \pi  M_P^2 r \sqrt{A B} \left(r F'+2 F\right)}{2 \ell+1}, \nonumber \\
&&a_3=\frac{2 \pi  \ell (\ell+1) M_P^2 \sqrt{\frac{A}{B}} \left(r F B'+2 B \left(r F'+F\right)\right)}{2 \ell r+r},~~~a_4=\frac{4 \pi  \ell (\ell+1) M_P^2 F \sqrt{A B}}{2 \ell+1}, \nonumber \\
&&a_5=-\frac{2 \pi  M_P^2 \left(r B A' \left(F-r F'\right)+A \left(F \left(r B'+2 B+2 \left(\ell^2+\ell-1\right)\right)+2 r B F'\right)\right)}{(2 \ell+1) F \sqrt{A B}}, \nonumber \\
&&a_6=\frac{2 \pi  M_P^2 r \sqrt{\frac{A}{B}} \left(r B'+4 B\right)}{2 \ell+1},~~~a_7=\frac{4 \pi  M_P^2 r^2 \sqrt{A B}}{2 \ell+1}, \nonumber \\
&&a_8=-\frac{16 \pi  \ell (\ell+1) M_P^2 \left(A \left(W' \left(2 r B'+\ell^2+\ell-2\right)+2 r B W''\right)-r B A' W'\right)}{(2 \ell+1) r^2 A}, \nonumber \\
&&a_9=-\frac{8 \pi  \ell (\ell+1) M_P^2 \left(r B A' W'-2 A \left(r B' W'+B \left(r W''-W'\right)\right)\right)}{(2 \ell+1) r A},~~~a_{10}=\frac{16 \pi  \ell (\ell+1) M_P^2 B W'}{2 \ell+1}, \nonumber \\
&&a_{11}=-\frac{8 \pi  \ell (\ell+1) M_P^2 \left(2 A \left(r B' W'+B \left(r W''+W'\right)\right)-r B A' W'\right)}{(2 \ell+1) r A},~~~a_{12}=-a_{10},~~~b_1=\frac{a_4}{2A},~~~b_2=-\frac{2 a_7}{A}, \nonumber \\
&&b_3=\frac{A'}{A^2}a_7,~~~b_4=-\frac{2a_2}{A},~~~b_5=-2b_1,~~~b_6=\frac{2a_{10}}{rA},~~~b_7=-\frac{r}{2}b_6,~~~b_8=\frac{\ell^2+\ell-2}{r^2}a_{10},~~~b_9=-\frac{2a_{12}b_1}{a_4}, \nonumber \\
&&c_1=\frac{\pi  M_P^2 \sqrt{\frac{B}{A}} \left(r A' \left(r F'+2 F\right)+2 A \left(2 r F'+F\right)\right)}{2 \ell+1},~~~c_2=-\frac{2 \pi  \ell (\ell+1) M_P^2 \sqrt{\frac{B}{A}} \left(r F A'+2 A \left(r F'+F\right)\right)}{2 \ell r+r}, \nonumber \\
&&c_3=-\frac{2 \pi  M_P^2 r \sqrt{\frac{B}{A}} \left(r A'+4 A\right)}{2 \ell+1},~~~c_4=\frac{2 \pi  M_P^2 \left(A \left(F \left(3 r B'+2 B+2 \left(\ell^2+\ell-1\right)\right)+2 r B F'\right)-r B A' \left(r F'+F\right)\right)}{(2 \ell+1) F \sqrt{A B}}, \nonumber \\
&&c_5=-\frac{8 \pi  \ell (\ell+1) M_P^2 B \left(r A'-2 A\right) W'}{(2 \ell+1) r A},~~~c_6=-\frac{2c_5}{r},~~~c_7=-c_5,~~~c_8=\frac{b_2}{2B},~~~d_1=-\frac{b_5}{2},~~~d_2=\frac{a_4}{r^2},~~~d_3=\frac{2a_4}{F}, \nonumber \\
&&d_4=-\frac{2a_4}{rF},~~d_5=-\frac{a_{10}}{A},~~d_6=\frac{2a_{10}}{rA},~~d_7=-\frac{2a_{10}}{r^2},~~d_8=\frac{16 \pi  \ell (\ell+1) M_P^2 B W' \left(2 \ell (\ell+1) A-\left(\ell^2+\ell-2\right) r A'\right)}{(2 \ell+1) r^3 A}, \nonumber \\
&&d_9=\frac{a_{10}}{A},~~~d_{10}=\frac{\ell (\ell+1)}{r^2}a_{10},~~~e_1=\frac{2 \pi  \ell (\ell+1) M_P^2 \left(A \left(F \left(r B'+2 B+\ell^2+\ell-2\right)+2 r B F'\right)-r B F A'\right)}{(2 \ell+1) r^2 A^{3/2} \sqrt{B}}, \nonumber \\
&&e_2=\frac{4b_1}{r},~~~e_3=\frac{16 \pi  \ell (\ell+1) M_P^2 \left(2 A \left(r B F'+2 (B-1) F\right)-r B A' \left(r F'+2 F\right)\right)}{(2 \ell+1) r^2 A F}, \nonumber \\
&&e_4=\frac{32 \pi  \ell (\ell+1) M_P^2 \left(r B A' \left(r^2 F'^2+r F F'+F^2\right)+A \left(-2 r^2 B F'^2+r (1-2 B) F F'-2 (B-1) F^2\right)\right)}{(2 \ell+1) r^3 A F^2}, \nonumber \\
&&e_5=-2b_1,~~~e_6=b_1,~~~e_7=b_1,~~~e_8=-\frac{(\ell-1) \ell (\ell+1) (\ell+2) F}{2 r^4} a_7,~~~e_9=e_3, \nonumber \\
&&f_1=\frac{2 \pi  M_P^2 r^2 W_s \sqrt{\frac{A}{B}}}{(2 \ell+1) \left(F_s^2-F_\lambda W_s \right)},~~~f_2=-\frac{2F_s}{W_s} f_1,~~~f_3=\frac{F_\lambda}{W_s} f_1.
\end{eqnarray}

\section{Expressions of $p_1,~p_2,\cdots$.}
\label{app-p}
Since most $p_1,\cdots$ have very long expressions, we assume that $F',~W',~F_\lambda,~F_s$ and $W_s$ are small quantities and truncate at the leading order. 
$p_1,~p_2$ and $p_3$ are exact.
\begin{eqnarray}
p_1&=&-\frac{32 \pi  \ell (\ell+1) M_P^2 W'^2}{(2 \ell+1) F \left(\frac{A}{B}\right)^{3/2}},~~~p_2=-\frac{2p_1}{r},~~~p_3=p_1, \nonumber \\
p_4&=&-\frac{32 \pi  \ell (\ell+1) M_P^2 W'}{(2 \ell+1) r^2 A^2 F \sqrt{\frac{A}{B}} \left(r B A'+A \left(-2
   B+\ell^2+\ell\right)\right)^2} \left(r A^2 B A' \left(W' \left(-3 \ell (\ell+1) \left(-2 r B'+\ell^2+\ell\right) \right.\right.\right. \nonumber \\
   &&\left.\left.\left.-4 B \left(3 r B'-8\right)-20 B^2\right)+8 r B
   \left(-2 B+\ell^2+\ell\right) W''\right)+r^2 A B^2 A'^2 \left(W' \left(3 r B'+14 B-6 \ell (\ell+1)\right)+4 r B W''\right) \right. \nonumber \\
   &&\left.-3 r^3 B^3 A'^3 W'+A^3 \left(3
   \ell^2 (\ell+1)^2 r B' W'+4 B^2 \left(W' \left(3 r B'+2 \left(\ell^2+\ell-4\right)\right)-4 \ell (\ell+1) r W''\right) \right.\right. \nonumber \\
   &&\left.\left.-2 \ell (\ell+1) B \left(W' \left(6 r B'+7 \ell^2+7 \ell-16\right)-2
   \ell (\ell+1) r W''\right)+8 B^3 \left(2 r W''+W'\right)\right)\right), \nonumber \\
p_5&=&\frac{2 \pi \ell (\ell+1) M_P^2 F}{(2 \ell+1) \sqrt{\frac{A}{B}}}, \nonumber \\
p_6&=&\frac{4 \pi  M_P^2 r^2 \sqrt{\frac{A}{B}}}{\ell \left(2 \ell^2+3 \ell+1\right) A F \left(r B A'+A \left(-2 B+\ell^2+\ell\right)\right)^2} \left(\left(2 \ell^2+2 \ell-1\right) r^2 B^2 A'^2 \right. \nonumber \\
&&\left.+2 \ell (\ell+1) r A B A' \left(-3 B+2 \ell^2+2 \ell-1\right)+\ell (\ell+1) A^2 \left(-6 \ell (\ell+1) B+6 B^2+\ell \left(2 \ell^3+4 \ell^2+\ell-1\right)\right)\right), \nonumber \\
p_7&=&\frac{8 \pi  \ell (\ell+1) \left(\ell^2+\ell-2\right) M_P^2 A^3 F}{(2 \ell+1) \left(\frac{A}{B}\right)^{3/2} \left(r B A'+A \left(-2 B+\ell^2+\ell\right)\right)^2},~~~p_8=\frac{32 \pi  \left(\ell^2+\ell-2\right) M_P^2 W' \left(r B A'+\ell (\ell+1) A\right)^2}{(2 \ell+1) A F \left(r B A'+A \left(-2 B+\ell^2+\ell\right)\right)^2}, \nonumber \\
p_9&=&-\frac{64 \pi \ell (\ell+1) \left(\ell^2+\ell-2\right) M_P^2 B W' \left(r B A'+\ell (\ell+1) A\right)}{(2 \ell r+r) \left(r B A'+A \left(-2 B+\ell^2+\ell\right)\right)^2}, \nonumber \\
p_{10}&=&-\frac{8 \pi  \left(\ell^2+\ell-2\right) M_P^2 r B \sqrt{\frac{A}{B}} \left(r B A'+\ell (\ell+1) A\right)}{(2 \ell+1) \left(r B A'+A \left(-2 B+\ell^2+\ell\right)\right)^2},~~~p_{11}=\frac{2 \pi  \ell (\ell+1) M_P^2 F}{(2 \ell+1) \sqrt{\frac{A}{B}}}, \nonumber \\
p_{12}&=&-\frac{4 \pi M_P^2 r^2 B \sqrt{\frac{A}{B}}}{\ell \left(2 \ell^2+3 \ell+1\right) F \left(r B A'+A \left(-2 B+\ell^2+\ell\right)\right)^2} \left(\left(2 \ell^2+2 \ell-1\right) r^2 B^2 A'^2 \right. \nonumber \\
&&\left.+2 \ell (\ell+1) r A B A' \left(-3 B+2 \ell^2+2 \ell-1\right)+\ell (\ell+1) A^2 \left(-6 \ell (\ell+1) B+6 B^2+\ell \left(2 \ell^3+4 \ell^2+\ell-1\right)\right)\right), \nonumber \\
p_{13}&=&-\frac{8 \pi  \ell (\ell+1) \left(\ell^2+\ell-2\right) M_P^2 A^5 F}{(2 \ell+1) \left(\frac{A}{B}\right)^{5/2} \left(r B A'+A \left(-2 B+\ell^2+\ell\right)\right)^2},~~~p_{14}=-\frac{32 \pi  \left(\ell^2+\ell-2\right) M_P^2 \epsilon  B W' \left(r B A'+\ell (\ell+1) A\right)^2}{(2 \ell+1) F \left(r B A'+A \left(-2 B+\ell^2+\ell\right)\right)^2}, \nonumber \\
p_{15}&=&\frac{64 \pi  \ell (\ell+1) \left(\ell^2+\ell-2\right) M_P^2 A B^2 W' \left(r B A'+\ell (\ell+1) A\right)}{(2 \ell r+r) \left(r B A'+A \left(-2 B+\ell^2+\ell\right)\right)^2}, \nonumber \\
p_{16}&=&\frac{8 \pi  \left(\ell^2+\ell-2\right) M_P^2 r A^3 \left(r B A'+\ell (\ell+1) A\right)}{(2 \ell+1) \left(\frac{A}{B}\right)^{3/2} \left(r B A'+A \left(-2 B+\ell^2+\ell\right)\right)^2},~~~p_{17}=-\frac{4 \pi  \ell (\ell+1) M_P^2 F}{(2 \ell+1) \sqrt{\frac{A}{B}}},~~~p_{18}=-\frac{8 \pi  \ell (\ell+1) M_P^2 F}{(2 \ell r+r) \sqrt{\frac{A}{B}}}, \nonumber \\
p_{19}&=&\frac{32 \pi  \ell (\ell+1) M_P^2 F_s \left(r B A'-2 A (B-1)\right)}{(2 \ell+1) r^2 A F_\lambda}, \nonumber \\
p_{20}&=&\frac{32 \pi  \ell (\ell+1) M_P^2 B}{(2 \ell+1) r^2 \left(r B A'+A \left(-2 B+\ell^2+\ell\right)\right)^2} \left(r A B A' W' \left(\left(\ell^2+\ell-2\right) r B'+4 \ell (\ell+1) B+\ell \left(\ell^3+2 \ell^2-5 \ell-6\right)\right) \right. \nonumber \\
&&\left.+\left(\ell^2+\ell-4\right) r^2
   B^2 A'^2 W'+A^2 \left(W' \left(\ell (\ell+1) \left(\left(\ell^2+\ell-2\right) r B'-2 \ell (\ell+1)\right) \right.\right.\right. \nonumber \\
   &&\left.\left.\left.+2 B \left(-2 \left(\ell^2+\ell-2\right) r B'+\ell^4+2 \ell^3+5 \ell^2+4 \ell-4\right)-4 \ell
   (\ell+1) B^2\right)-4 \left(\ell^2+\ell-2\right) r B^2 W''\right)\right), \nonumber \\
p_{21}&=&\frac{32 \pi \ell (\ell+1) M_P^2 F_s \left(2 A (B-1)-r B A'\right)}{(2 \ell+1) r^2 A F_\lambda},~~~p_{22}=\frac{64 \pi  \ell (\ell+1) M_P^2 B W'}{(2 \ell+1) r^2}, \nonumber \\
p_{23}&=&\frac{8 \pi  M_P^2 B \sqrt{\frac{A}{B}}}{(2 \ell+1) \left(r B A'+A \left(-2 B+\ell^2+\ell\right)\right)^2} \left(r A B A' \left(\left(\ell^2+\ell+1\right) r B'+\left(4 \ell^2+4 \ell-2\right) B-\ell \left(2 \ell^3+4 \ell^2+\ell-1\right)\right) \right.\nonumber \\
   &&\left.+\left(-2 \ell^2-2 \ell+1\right) r^2 B^2 A'^2+A^2 \left(B \left(\left(\ell^2+\ell-2\right)^2-2 \left(\ell^2+\ell+1\right) r B'\right) \right.\right.\nonumber \\
   &&\left.\left.+\ell \left(\ell^3+2 \ell^2+2 \ell+1\right) r
   B'+2 \left(\ell^2+\ell-2\right) B^2\right)\right),~~~p_{24}=\frac{2 \pi  \ell^2 (\ell+1)^2 M_P^2}{(2 \ell+1) r^2},
\end{eqnarray}

\begin{eqnarray}
p_{25}&=&\frac{64 \pi  \ell (\ell+1) M_P^2 F_s F'}{(2 \ell+1) r^2 F_\lambda }, \nonumber \\
p_{26}&=&\frac{64 \pi  \ell (\ell+1) M_P^2 \left(\left(\ell^4+2 \ell^3+\ell^2-4\right) W'-2 r \left(r W^{(3)}-2 W''\right)\right)}{\left(2 \ell^3+3 \ell^2-3 \ell-2\right) r^3},~~~p_{27}=-\frac{2 \pi  \ell \left(\ell^3+2 \ell^2-\ell-2\right) M_P^2}{(2 \ell+1) r^2}, \nonumber \\
p_{28}&=&-\frac{2 \pi  M_P^2 r^2}{(2 \ell+1) F_\lambda}, \nonumber \\
p_{29}&=&\frac{8 \pi \left(\ell^2+\ell-2\right) M_P^2}{(2 \ell r+r) \left(-3 A+\ell^2+\ell+1\right)^3} \left(6 \left(\ell^2+\ell+1\right) A^3-4 \left(\ell^2+\ell+1\right)^2 A^2 \right.\nonumber \\
&&\left.+\left(\ell^6+3 \ell^5+6 \ell^4+7 \ell^3+7 \ell^2+4 \ell+2\right) A-3 A^4-\left(\ell^2+\ell+1\right)^2\right), \nonumber \\
p_{30}&=&-\frac{8 \pi  \ell \left(\ell^3+2 \ell^2-\ell-2\right) M_P^2 A \left(-3 \left(\ell^2+\ell+1\right) A^2+\left(\ell^2+\ell+1\right)^2 A+3 A^3-\ell^2-\ell-1\right)}{(2 \ell+1) r^2 \left(-3 A+\ell^2+\ell+1\right)^3}. \nonumber
\end{eqnarray}

\section{Expressions of $s_3$.}
\label{app-s}
$s_3$ is given by
\begin{eqnarray}
s_3=\frac{z_1}{z_2}, \nonumber
\end{eqnarray}
where $z_1$ and $z_2$ are given by
\begin{eqnarray}
z_1&=&4 \pi M_P^2 \left(64 r^3 A^2 F_s^2 F' \left(r B A' \left(r F'+2 F\right)-2 A \left(r B F'+2
   (B-1) F\right)\right)^2 \right. \nonumber \\
   &&\left.-\frac{1}{F^2} \left(-r^2 A F_\lambda A' \left(3 F^2 B'+2 B F F'+4 r B
   F'^2\right)+3 r^2 B F^2 F_\lambda A'^2-2 A^2 \left(F^2 \left(3 F_\lambda \left(r^2
   B''-2 B+2\right)+r^3 F'\right) \right.\right.\right. \nonumber \\
   &&\left.\left.\left.-4 r^2 B F_\lambda F'^2-2 r (B-1) F F_\lambda
   F'\right)\right) \left(-32 r^2 B^2 A'^2 \left(r F'+2 F\right)^2 \left(F_s^2-F_\lambda
   W_s\right) \right.\right. \nonumber \\
   &&\left.\left.+128 r A B A' \left(r F'+2 F\right) \left(r B F'+2 (B-1) F\right)
   \left(F_s^2-F_\lambda W_s\right)+A^2 \left(-128 r^2 B^2 F'^2
   \left(F_s^2-F_\lambda W_s\right) \right.\right.\right. \nonumber \\
   &&\left.\left.\left.-512 r (B-1) B F F' \left(F_s^2-F_\lambda  W_s\right)-512 (B-1)^2 F^2 \left(F_s^2-F_\lambda W_s\right)+r^6 F^3
   F_\lambda\right)\right)\right), \nonumber \\
z_2&=&3 r^6 A^2 F_\lambda \sqrt{\frac{A}{B}} \left(r^2 A F_\lambda A' \left(3 F^2 B'+2 B F F'+4 r B F'^2\right)-3 r^2 B F^2 F_\lambda A'^2
 \right. \nonumber \\
 &&\left.+2 A^2 \left(F^2 \left(3 F_\lambda \left(r^2 B''-2 B+2\right)+r^3 F'\right)-4 r^2 B F_\lambda F'^2-2 r (B-1) F F_\lambda F'\right)\right). \nonumber   
\end{eqnarray}
In particular, for $f(R)$ theories in which $F_s=W_s=0$, $s_3$ is drastically simplified to
\begin{equation}
s_3=\frac{4 \pi M_P^2 F}{3 \sqrt{\frac{A}{B}}}. 
\end{equation}

\end{document}